\documentclass[aip,reprint,amsmath,amssymb,twocolumn]{revtex4-1}
\usepackage{graphicx}% Include figure files
\usepackage{dcolumn}% Align table columns on decimal point
\usepackage{bm}% bold math

\usepackage{textcomp}
\newcommand{\mic}{~\mu \rm m}
\draft % marks overfull lines with a black rule on the right

\begin{document}

\title{Multiple-octave spanning mid-IR supercontinuum generation in bulk quadratic nonlinear crystals} %Title of paper

\author{Binbin Zhou}
%\email[]{zhou@fotonik.dtu.dk}
%\thanks{}
\affiliation{DTU Fotonik, Department of Photonics Engineering, Technical University of Denmark, DK-2800 Kgs. Lyngby, Denmark}
\author{Morten Bache}
\email[]{moba@fotonik.dtu.dk}
\homepage[]{www.fotonik.dtu.dk/uno}
%\thanks{}
\affiliation{DTU Fotonik, Department of Photonics Engineering, Technical University of Denmark, DK-2800 Kgs. Lyngby, Denmark}

\date{\today}

\begin{abstract}
    Bright and broadband coherent mid-IR radiation is important for exciting and probing molecular vibrations. Using cascaded nonlinearities in conventional quadratic nonlinear crystal like lithium niobate, self-defocusing near-IR solitons have been demonstrated that led to very broadband supercontinuum generation in the visible, near-IR and short-wavelength mid-IR. Here we conduct an experiment where a mid-IR crystal pumped in the mid-IR gives multiple-octave spanning supercontinua. The crystal is cut for noncritical interaction, so the three-wave mixing of a single mid-IR femtosecond pump source leads to highly phase-mismatched second-harmonic generation. This self-acting cascaded process leads to the formation of a self-defocusing soliton at the mid-IR pump wavelength and after the self-compression point multiple octave-spanning supercontinua are observed (covering 1.6-$7.0\mic$). The results were recorded in a commercially available crystal LiInS$_2$ pumped in the 3-$4\mic$ range, but other mid-IR crystals can readily be used as well.
\end{abstract}

\pacs{}% insert suggested PACS numbers in braces on next line

\maketitle %\maketitle must follow title, authors, abstract and \pacs

\section{Introduction}

%If instead near-IR laser pulses are used, overtones of the vibrations are excited, reducing the desired control.
High-power supercontinuum (SC) sources\cite{alfano:2006} are laser pulses that have achieved octave-spanning broadened spectra by passing the laser through a bulk nonlinear medium. They are widely used in spectroscopy as broadband "white-light" probes that in a single shot can assess a wide range of frequencies. The mid-infrared (mid-IR) range, spanning from 3.0-$20.0\mic$ holds a tremendous scientific and technological potential. It contains the fundamental frequencies of vibrational stretching modes of the important C-H, O-H and N-H bonds that lie in the 2.5-$3.5 \mic$  wavelength range, and the carbon doublet and triplets in the 4-$7\mic$ range. From 7-$20\mic$ lies the so-called “fingerprint region” where all organic compounds have a unique spectral absorption pattern due to single-bond bending modes. With a broadband mid-IR SC source the vibrational modes can be probed with femtosecond resolution\cite{Calabrese2012}.

Unlike most fiber-based SC sources, which rely on soliton formation\cite{dudley:2006}, the bulk SC source relies mainly on broadening induced  by self-phase modulation (SPM). At high powers a filament is created by the intense laser pulse. This happens because the beam experiences focusing to a small spot by the nonlinear self-focusing Kerr-lens effect that arises because the Kerr nonlinear refractive index is positive. Filamentation is often undesirable as it can lead to a spatially incoherent beam. However, under controlled circumstances a so-called white-light continuum (WLC) can be generated. In a WLC typically a single filament is created ensuring a good spatial coherence. Both the SPM-induced broadening and the WLC techniques are widely used in the near-IR but little has been done in the mid-IR: Broadband continua were found when pumping with mid-IR pulses through SPM alone (i.e. before filamentation set in)\cite{Corkum:1985,Ashihara:2009-ol,Pigeon2014}, which is limited in energy by onset of small-scale filamentation, or by generating a WLC by an increased peak power\cite{silva:2012,Kartashov:2012b,Darginavicius2013,Liang2015,Mouawad2015}. Alternatively one may start with a femtosecond near-IR source and mix the frequency-converted harmonics in air and use four-wave mixing to achieve broadband mid-IR radiation\cite{Fuji:2007,Petersen:2010}, but with a very low yield. %With similar techniques near-IR laser pulses can be converted to few-cycle or even single-cycle duration to practically any desired (longer) wavelengths, but again the yield is very low \cite{Calabrese2012}.

Finally, supercontinuum generation has also been observed in bulk quadratic nonlinear crystals \cite{Fuji:2005,Srinivas:2005,Kumar:2007,Krupa2015} and waveguides \cite{Langrock:2007,Phillips:2011-ol,Guo-LN-exp-2015}. Around phase-matching the pump spectral broadening is accompanied by second-harmonic generation, which adds to the spectral bandwidth. However, even when tuning away from SHG phase matching ($\Delta k=k_2-2k_1\neq 0$) the spectral broadening can become surprisingly large. %; well over an octave of bandwidth has been observed in waveguides \cite{Langrock:2007,Phillips:2011-ol,Guo-LN-exp-2015}.
This has two likely explanations: 1) the SPM initiated continuum around the pump can mix through sum-frequency generation (SFG) with the pump. 2) The phase-mismatched SHG process generates a "cascaded" SPM-like cubic nonlinearity $n_{2, \rm casc}\propto -d_{\rm eff}^2/\Delta k$. This contributes to the material Kerr SPM nonlinearity, $n_{2,\rm Kerr}$, which is self-focusing, to give an overall effective nonlinearity $n_{2,\rm eff}=n_{2, \rm casc}+n_{2, \rm Kerr}$. For $\Delta k<0$ the cascaded effect is also self-focusing, so the supercontinuum is in bulk often accompanied by filamentation. 
%As an additional where the high-frequency part of the spectrum mainly accompanied by filamentation.

We recently \cite{zhou:2012} showed that for $\Delta k>0$ i.e. when the cascaded effect is self-defocusing, the $n_{2,\rm eff}>0$ condition can be met so the supercontinuum is filamentation free, and the spectral broadening is very large because it is possible to excite solitons when pumping in the normal dispersion regime. An octave-spanning spectrum was recorded in bulk lithium niobate (LN) by pumping well below its zero-dispersion wavelength (ZDW), and akin to the filamentation approach almost all of the energy is a transferred into the supercontinuum. %, and subsequently we showed that the few-cycle near-IR soliton can become phase matched to a mid-IR resonant radiation peak (also known as dispersive waves) \cite{Zhou:2015}, located beyond the ZDW. Similar results were found in BBO \cite{Zhou:2015-OL}. %Thus, quadratic nonlinear crystals are excellent candidates for supercontinuum generation in bulk media with near-IR pump wavelengths.

LN is particularly interesting because it exploits "noncritical" $ee\rightarrow e$ three-wave mixing to achieve cascading through the large diagonal quadratic nonlinear tensor component ($d_{33}\simeq 25$ pm/V), and this has the additional advantage of zero spatial walk-off to the second-harmonic (SH). 
Here we show that multiple-octave spanning mid-IR supercontinua can be generated in a crystal similar to LN but transparent in most of the mid-IR. 
We recently showed that many LN-like crystals exist in the mid-IR, i.e. with large quadratic tensor components and noncritical interaction, that can be exploited in much the same way except a mid-IR femtosecond pump laser must be used \cite{Bache:2013-midIR}. The present paper is the first experimental realization of these predictions, and we chose to study the crystal lithium thioindate (LiInS$_2$, LIS). We show that when pumped with $\sim 50~\mu \rm J$ sub-100 fs pulses from $3.0-3.9\mic$ around its ZDW, LIS gives supercontinua spanning well over 2 octaves. Importantly the spectra are broad enough to cover the range 2.5-$6.0\mic$ in a single pulse, which is important to probe fundamental vibrations of OH, CH, NH, OD, metallic CO and organic CO bonds, all in a single pulse. 

\section{Background}

Cascaded nonlinearities were noted long ago \cite{Ostrovskii:1967}, where a basic theoretical treatment predicted that under strongly phase-mismatched parametric interaction, the pump pulse may experience self-action, i.e. self-phase modulation of the spectrum. This is because for the simplest case of second-harmonic generation (SHG), the strong phase mismatch $|\Delta k|\gg 0$ implies that the pump at $\omega_1$ will be partially upconverted to the second harmonic (SH) at $\omega_2=2\omega_1$ within a coherence length $l_{\rm coh}=|\Delta k|/\pi$, but after another coherence length the SH photons will be back converted to the pump. Under the strong phase-mismatch limit ($|\Delta k|L\gg 2\pi$), $l_{\rm coh}$ is much shorter than the crystal length $L$, so this up- and down-conversion process happens multiple times: this is the reason it is called a cascaded nonlinearity. The cascaded nonlinearity was first experimentally verified in \cite{desalvo:1992}, where the cascaded nonlinearity was shown to be tunable in sign and strength $n_{2,\rm casc}\propto -d_{\rm eff}^2/\Delta k$, and where importantly it became clear that a self-defocusing effect was accessible ($n_{2,\rm casc}<0$, requiring $\Delta k>0$). 

A controllable leading-order self-defocusing nonlinearity is quite unique, and quite some effort was therefore invested in applications of this effect in bulk crystals. One application was pulse compression of energetic pulses, because in bulk glasses the required spectral broadening was accompanied by self-focusing, limiting the pulse energy. The idea was to compensate for the material Kerr self-focusing, making $n_{2,\rm eff}<0$ by using cascaded effects, and this would give a pulse compressor without filamentation limitations; it was first investigated in \cite{liu:1999}, where an SPM-induced self-defocusing spectral broadening was induced in a quadratic nonliner crystal (BBO, beta-barium borate). The effective self-defocusing nonlinearity gave a negative chirp across the pulse, so the pulse could be compressed by passing it through a piece of bulk glass with positive normal dispersion. 

A similar experiment was conducted by Ashihara et al. \cite{ashihara:2002}, but instead of compressing the pulse "externally", they achieved soliton self-compression inside the crystal. This is because at the pump wavelength (800 nm from a TiSa amplifier) BBO has normal group-velocity dispersion (GVD), and for $\beta_2>0$ exciting a soliton requires  $n_{2,\rm eff}<0$. Although the soliton excitation was confirmed, the compression ratio was moderate. This is because BBO has a large group-velocity mismatch (GVM) between pump and SH at this wavelength, and indeed later experiments at longer wavelengths have shown few-cycle soliton self-compression in BBO due to a reduced GVM \cite{moses:2006,Zhou:2015-OL}. 

Once the soliton has formed, it may become phase-matched to generate resonant radiation (i.e. a dispersive wave) \cite{Skryabin:2010}. Since the self-defocusing soliton needs normal dispersion to form, the dispersive wave will naturally be generated in the anomalous dispersion regime, i.e. to the long-wavelength side \cite{bache:2010e,bache:2011a}. This was recently experimentally verified in bulk BBO and LN \cite{Zhou:2015-OL,Zhou:2015}, and together with the soliton the dispersive wave(s) constitute the octave-spanning supercontinuum in the self-defocusing soliton case, which for LN can extend from 1.0-$4.0\mic$. 

This type of dispersive wave is quite similar to the one observed in Kerr media, since it is mediated by four-wave mixing (FWM, degenerate \cite{Zhou:2015-OL,Zhou:2015} or non-degenerate case \cite{Zhou:2015-OL,Liu2015}), but with the cascaded quadratic nonlinearity we also found that there is a unique three-wave mixing (TWM) analogue. This was recently observed experimentally for the first time \cite{zhou:PhysRevA.90.013823}, where we found that in BBO the SH beam had a parametrically tunable peak, i.e. tunable by fixing the pumping conditions and changing only a single parameter (the phase-mismatch parameter). It was tunable across the entire visible range and was excited due to a dispersive wave phase-matching condition to the pump near-IR soliton, but unlike the Kerr case it relies on direct TWM, which is exactly why it is parametrically tunable. Its phase-matching condition is related to the so-called nonlocal effect observed in cascaded nonlinearities \cite{bache:2007a,bache:2008}, which is a consequence of soliton-dispersive wave intra-dispersion effects (phase-mismatch and GVM). Such a tunable dispersive wave is also not common in Kerr media (only in gas-filled fibers has this been demonstrated \cite{Mak2013}).

In BBO, the SHG phase-matching scheme is "critical" or birefringence phase-matching, i.e. so-called type I where the phase-matching is angle-dependent, and as a consequence the pump and SH have orthogonal polarization (in BBO the $oo\rightarrow e$ scheme is used). Other soliton self-compression experiments used such a pumping configuration in BBO or LN \cite{moses:2007,ashihara:2004,Zeng:2008}; the main advantage is that one can find the SHG phase-matching condition, and the slowly tune away from it by rotating the crystal to go to the cascading limit; this gives huge cascaded nonlinearities because as soon as cascading sets in, it scales as $1/\Delta k$ and the maximum is found around $\Delta kL=\pi$. The angle tuning onset of cascading is also a very accurate way of determining the $n_{2,\rm Kerr}$ value \cite{moses:2006b,Bache:2013}, as it is often possible to find the tuning angle where $n_{2,\rm eff}=0$. The disadvantages of type I interaction are that $d_{\rm eff}$ values are moderate and that the pump and SH experience spatial walk off. Thus, the crystal length is limited by this constraint (or, equivalently, large enough pump spot sizes must be used to minimize spatial walk-off effects). 

An entirely different approach is to use "non-critical" interaction, i.e. so-called type 0 where pump and SH have the same polarization and the phase-matching conditions do not depend very much on angle (hence the name non-critical). The advantages are that one may exploit the large diagonal tensor components, e.g. the $d_{33}$ of LN, and that spatial walk-off is nil. For supercontinuum generation the added bonus is that the SH has the same polarization as the pump, so there can be a considerable harmonic extension of the continuum (multiple octaves have been observed in LN waveguides \cite{Langrock:2007,Phillips:2011-ol}). The disadvantages are that the intra-harmonic dispersion (i.e. between pump and SH) is very large (so $\Delta k$ and GVM are very big), and that phase-matching is impossible (impossible to get $\Delta k=0$ and locate the SHG phase matching point as well as the maximum cascading point $\Delta kL=\pi$). Despite the large $\Delta k$, the cascading nonlinear index $n_{2,\rm casc}=-d_{\rm eff}^2/\Delta k$ can be quite significant: this is because $d_{\rm eff}$ is often very large and is thereby able to compensate for the large $\Delta k$. Importantly, for type 0 interaction $\Delta k>0$ if pumped not too close to an UV absorption region, because the SH will always have a larger refractive index than the pump (remember $\Delta k\propto [n(\omega_2)-n(\omega_1)]$). This means that in type 0 the cascaded SHG nonlinearity in a bulk crystal will always be self-defocusing. The question is merely: is it strong enough to overcome the material Kerr nonlinearity and generate an effective self-defocusing effect? We can pose this limit as a figure-of-merit parameter ${\rm FOM}=|n_{2,\rm casc}|/n_{2,\rm Kerr}$ and if ${\rm FOM}>1$ then the self-defocusing nonlinearity is dominating. LN has an ${\rm FOM}>1$ in the near-IR, as we used this to excite a temporal few-cycle soliton\cite{zhou:2012} when pumping in the normal GVD regime below its ZDW of $1.92\mic$, followed by octave-spanning supercontinuum generation \cite{Zhou:2015}. %Recently the solitonic nature of the compressed pulse was verified by observing the so-called soliton-induced resonant radiation (a.k.a. Cherenkov radiation, or dispersive wave formation) in the mid-IR\cite{Zhou:2015}, which only happens when the near-IR pulse is a soliton \cite{Skryabin:2010}. 

A similar approach to what we attempt here was conducted recently by Ashihara et al.\cite{Ashihara:2009-ol}, where GaAs was pumped in the mid-IR just below its ZDW to give significant spectral broadening (well over 2000 nm of bandwidth) at $5.0\mic$. GaAs has a very large quadratic nonlinearity, and is as such a good cascading candidate, but we calculate its FOM to be less than unity, i.e. self-focusing is dominating, and this was also noted by the authors. Therefore the spectral broadening came purely from self-focusing SPM due to the high material $n_{2,\rm Kerr}$ of GaAs. Finally, we mention that in quadratic nonlinear crystals it is also possible to perform adiabatic near-IR to mid-IR frequency conversion and achieve octave-spanning bandwidths \cite{Suchowski2013}.

\section{Lithium thioindate}

Using LN as an inspiration, we recently calculated the FOM for a range of mid-IR crystals\cite{Bache:2013-midIR}, who all have very large diagonal tensor components and possibility of getting crystal cut for non-critical SHG. One of the main crystals that attracted our interest was LIS, because it has a ZDW around $3.53\mic$ and an ${\rm FOM}\simeq 2$. Furthermore it is commercially available in quite big samples. Our simulations showed soliton formation and resonant radiation in the mid-IR, eventually giving supercontinuum radiation over several octaves in the mid-IR. 

\begin{figure}[tb]
\includegraphics[width=0.49\linewidth]{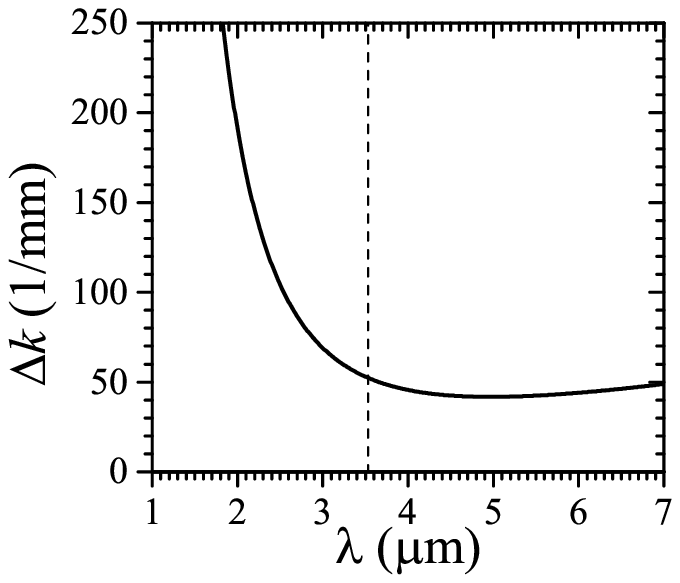}
\includegraphics[width=0.49\linewidth]{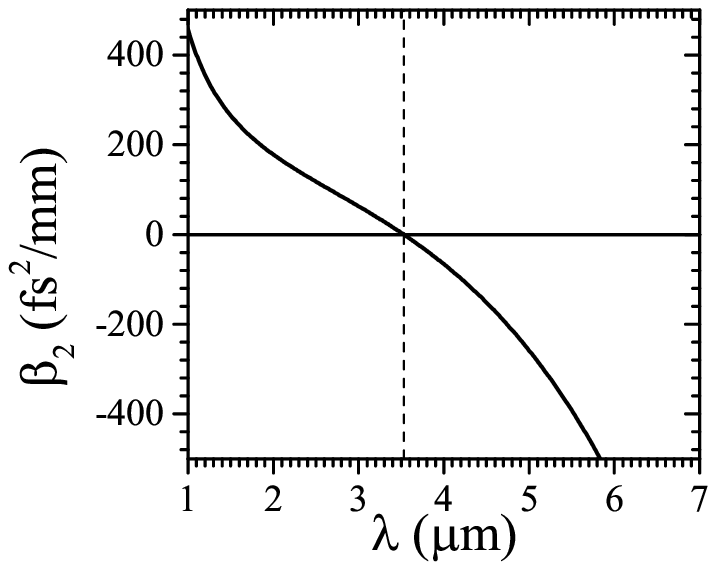}\\
\includegraphics[width=0.9\linewidth]{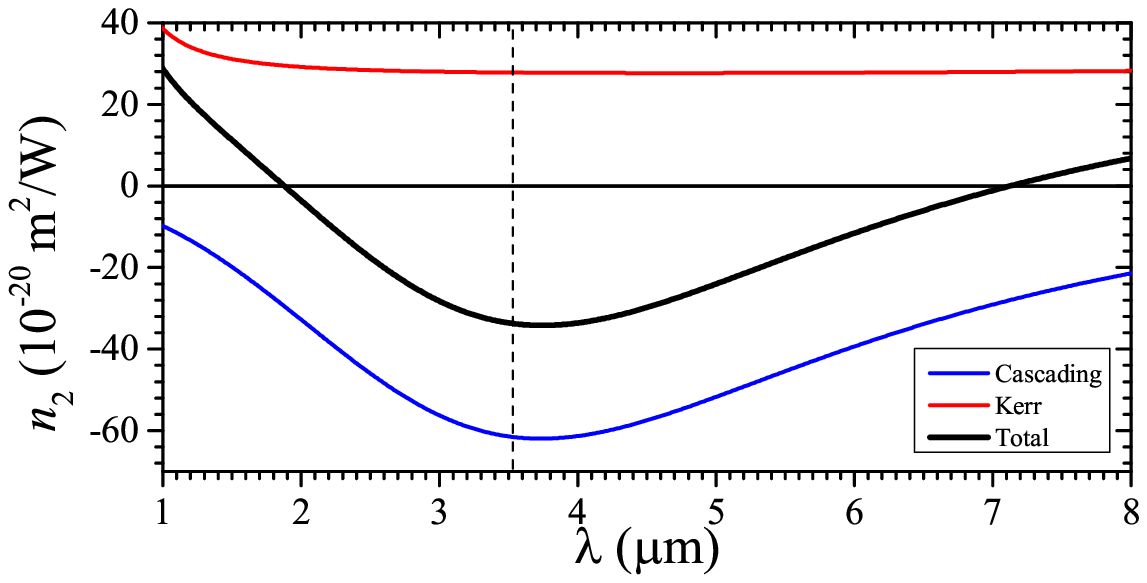}
\caption{\label{fig:n2} Top: phase-mismatch and GVD parameter; bottom: calculated cubic nonlinear refractive indices in LIS cut for interaction in the $XY$ plane ($\theta=\pi/2$ and $\phi=0$); thus the cascading channel is $ss\rightarrow s$, where the pump polarization is along the slow index $n_Z$, and the Kerr SPM nonlinearity is calculated for the $sss\rightarrow s$ interaction. The vertical dashed line denotes the ZDW.}%
\end{figure}

\begin{figure*}[t]
\includegraphics[width=0.45\linewidth]{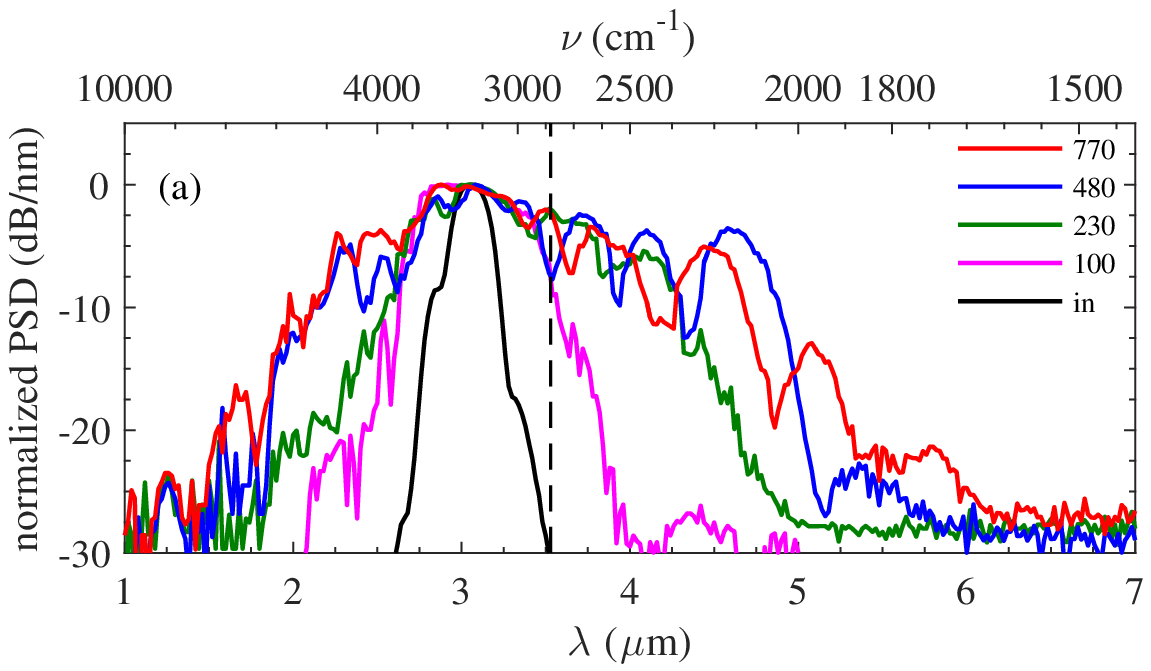}
\includegraphics[width=0.45\linewidth]{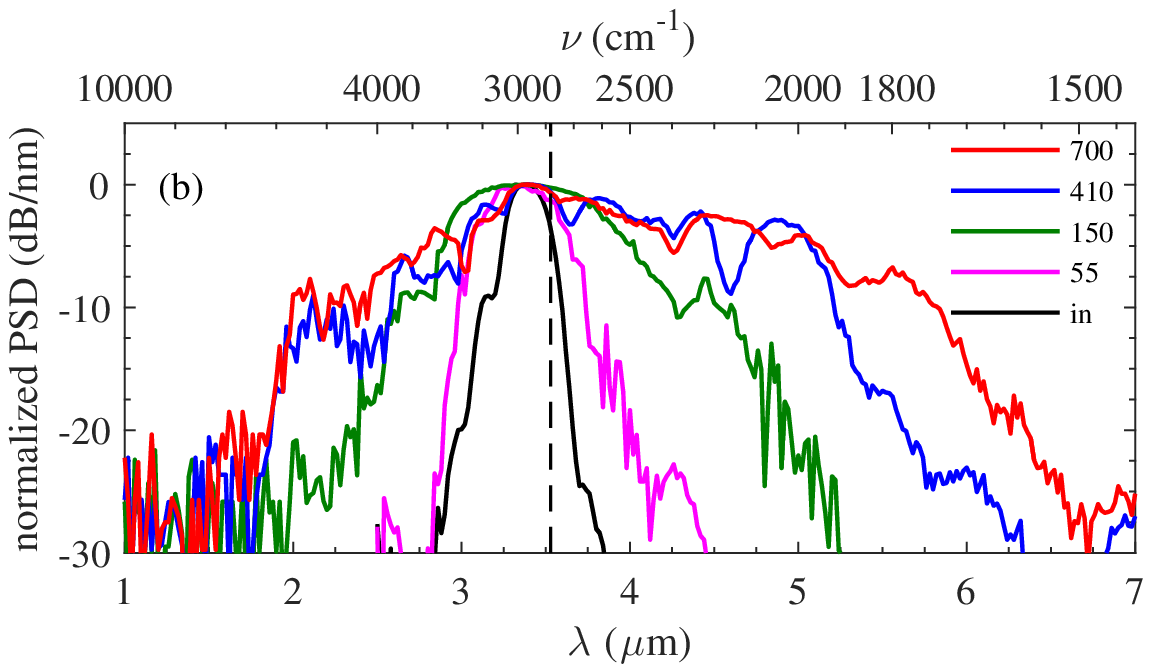}\\
\includegraphics[width=0.45\linewidth]{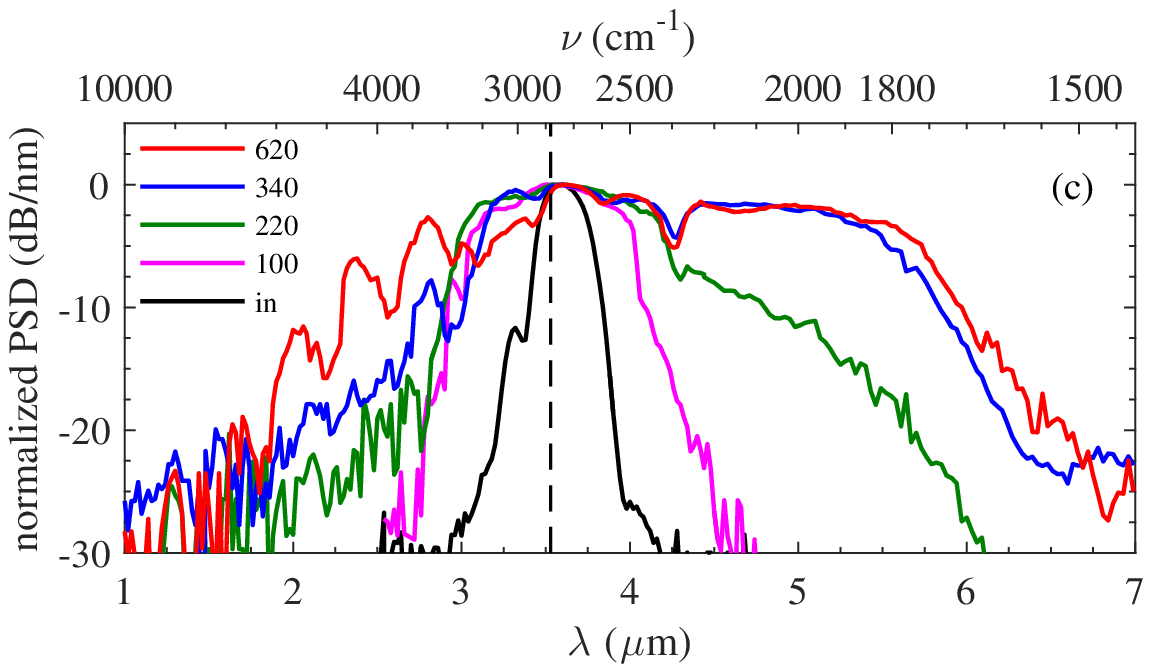}
\includegraphics[width=0.45\linewidth]{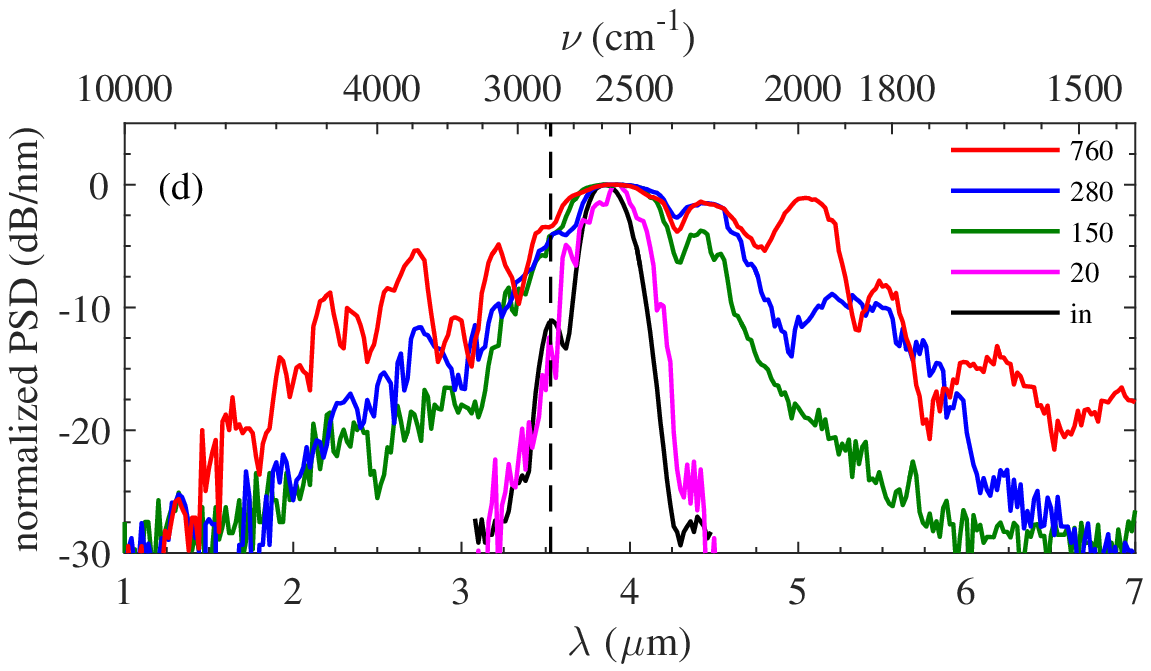}%
\caption{\label{fig:SCG} Experimentally recorded supercontinua in 15 mm LIS for pump wavelengths of (a) $\lambda=3.05\mic$, (b) $3.39\mic$, (c) $3.60\mic$ and (d) $3.86\mic$, and for various pump intensities (indicated in $\rm GW/cm^2$). The power spectral density (PSD) normalized to its peak value is shown. The vertical dashed line denotes the ZDW. The top axis shows the frequency in inverse wavenumbers $\nu=1/\lambda$, as often used in spectroscopy. Note the absorption dip at the CO$_2$ line ($\nu\simeq 2,600~\rm cm^{-1}$).}%
 \end{figure*}

For the present experiment we used a commercially available LIS crystal (Moltech), 15 mm long and $6\times6~\rm mm^2$ aperture. LIS belongs to the biaxial $mm2$ point group and our sample was cut with $\theta=90^\circ$ and $\phi=0$. This gives a maximum quadratic nonlinearity $d_{\rm eff}=d_{33}$ where $d_{33}=16$ pm/V@$2.3 \mic$ \cite{Fossier:2004}. This cut makes the SHG noncritical so both pump and SH have the same polarization (both are "slow" waves, with linear refractive index given by $n_Z$, see more details in \cite{Bache:2013-midIR}). The SHG process is highly phase-mismatched (coherence length $l_{\rm coh}=\pi/\Delta k \simeq 60\mic$), but notably not as much as in LN (where the coherence length can be an order of magnitude shorter). Due to the high quadratic nonlinearity the cascading Kerr nonlinear index is large ($n_{2,\rm casc}=-2\omega_1 d_{\rm eff}^2/[\varepsilon_0 c^2 n^2(\omega_1)n(\omega_2)\Delta k]\simeq -60\cdot 10^{-20}~\rm m^2/W$ in the pump range of interest). The competes with the material Kerr nonlinearity, and for LIS the cubic nonlinear tensor components are not known. However, from its rather high bandgap along the $Z$ direction ($E_g=3.55$ eV \cite{Fossier:2004}) the two-band model (see \cite{Bache:2013-midIR} for details) estimates the Kerr nonlinearity to be around $+30\cdot 10^{-20}~\rm m^2/W$. We therefore expect the overall cubic nonlinearity to be self-defocusing. This is summarized in Fig. \ref{fig:n2}, which shows the wavelength dependence of the material and cascaded Kerr nonlinear indices, calculated as discussed in \cite{Bache:2013-midIR}. In the range 2-$7\mic$ the effective nonlinearity is predicted to be self-defocusing, and the deducted FOM (i.e. the ratio between the cascaded and Kerr nonlinearity) is as high as 2, very similar to LN.

\section{Results}

We pumped the LIS crystal with 85 fs 50 $\mu$J pulses in the 3.0-$3.9\mic$ range from a commercial 1 kHz TiSa-amplifier based optical parametric amplifier (OPA) followed by a non-collinear difference-frequency generation stage. The pulses had around 200-250 nm bandwidth and were close to transform limit. The input beam was loosely focused (0.27 mm FWHM). The intensity controlled by adjusting the pump power level of the OPA. The output was measured with an FPAS-1600 spectrometer with a cooled MCT detector from Infrared Systems, and long-pass filters were used to selectively cover the 1-7$\mic$ range.

The results for different pump intensities and pump wavelengths are summarized in Fig. \ref{fig:SCG}. Generally, we needed around $150~\rm GW/cm^2$ to see an octave-spanning supercontinuum. The bandwidths in the maximum peak power cases exceed 2 octaves (1.6-$7.0 \mic@-20$ dB for $\lambda=3.86\mic$ pump). For the two cases that used pumping close to the ZDW, the supercontinuum is very flat across the central range, while in the two cases pumping further away the supercontinuum is more modulated.

The broadest supercontinuum was observed when pumping above the ZDW. This is surprising as we would have expected the broadest supercontinuum in the soliton case, i.e. when pumping below the ZDW. The soliton itself is very broadband (typically close to single cycle) and is furthermore also able to excite the resonant radiation in the anomalous dispersion regime above the ZDW, which further extends the bandwidth. We believe the explanation lies in the proximity of the pump wavelength to the ZDW, which means that even when pumping above the ZDW the early-stage spectral broadening leaks into the normal dispersion regime. When the most intense pulses were applied this allows a soliton to form despite pumping in the non-solitonic regime (and as such it is kind of the reverse case recently found in \cite{bache:2010e,Zhou:2015-OL}).

Numerical simulations were used to understand the experimental data. These were plane-wave simulations using the so-called Nonlinear Analytical Envelope Equation \cite{conforti:2010PRA,Conforti2013}, which is an envelope-like approach that is actually modeling the carrier wave, and has therefore sub-carrier wave resolution and includes a complete expansion of the nonlinear terms. The NAEE was recently extended to include both quadratic, cubic and delayed Raman effects \cite{bache:2016-NAEE}. The Raman mode was taken the same as in \cite{Bache:2013-midIR}, and the best agreements with the experimental results were seen using $n_{2,\rm Kerr}=50\cdot 10^{-20}~\rm m^2/W$ and a Raman fraction $f_R=0.2$; this implies that the electronic Kerr nonlinearity is $40\cdot 10^{-20}~\rm m^2/W$, quite close to the value predicted by the two-band model in Fig. \ref{fig:n2}. Finally, for the selected cut, LIS has no SHG coupling to the orthogonal polarization (for the $ss\rightarrow f$ channel $d_{\rm eff}=0$). As the cubic tensor components are taken isotropic, they do not have any cross-polarization coupling either, so we only need to model the input $s$ polarization.

\begin{figure}[tb]
\includegraphics[width=\linewidth]{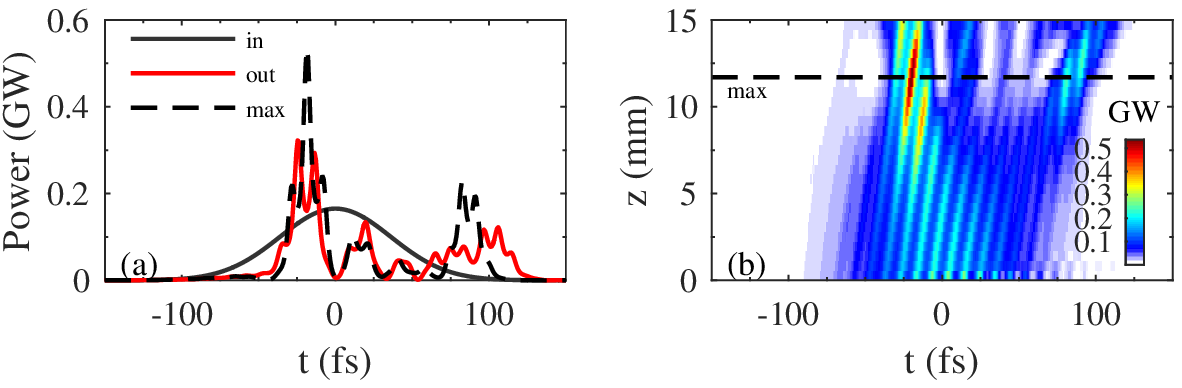}
\includegraphics[width=\linewidth]{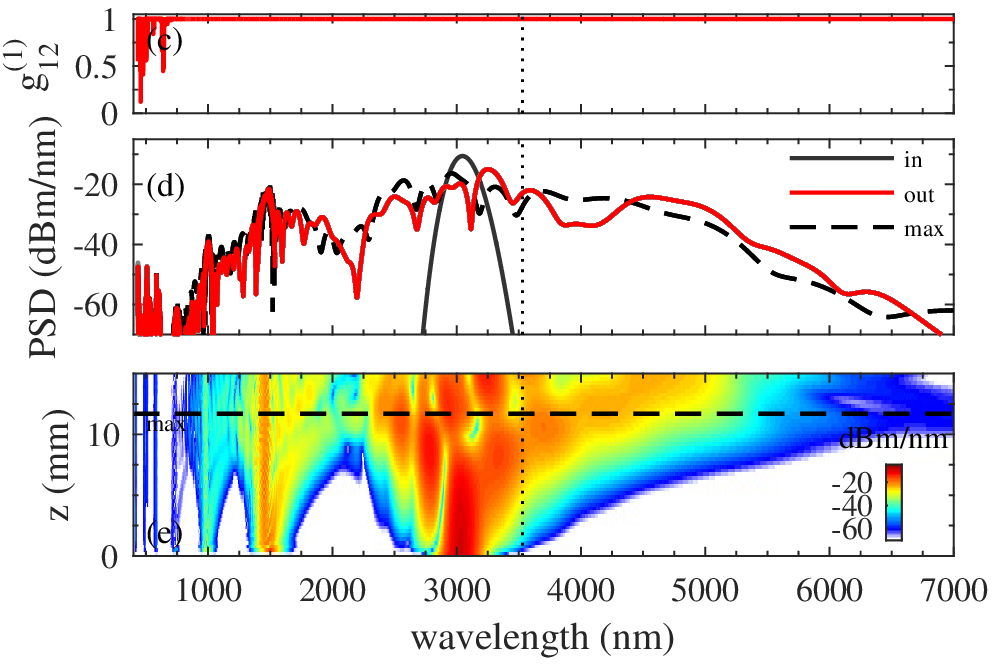}
\caption{\label{fig:sim} 
Numerical simulation at $\lambda=3.05\mic$ with 200 GW/cm$^2$ peak input intensity. (a) and (b) show the temporal evolution and (d) and (e) the spectral evolution of a single noise realization. In each case the "max" case shows the cut for maximum peak power in time domain, i.e. where the soliton self-compresses after 11 mm propagation. The output PSD in (d) and the coherence function $g_{12}^{(1)}$ in (e) were averaged over 20 noise realizations. The modulations on the envelope in (a) and (b) are caused by  harmonic generation along the polarization direction of the pump since this is included in the NAEE model. 
}%
\end{figure}

An specific simulation example is shown in Fig. \ref{fig:sim}, where we chose $\lambda=3.05\mic$ and a moderate intensity, corresponding to the green curve in Fig. \ref{fig:SCG}(a). The time domain in (a) and (b) show that a self-compressed soliton forms after 11 mm, but it is accompanied with two minor satellite pulses. Such pulse splitting may be caused by competing nonlinearities to the SPM from a combination of the Raman effect and self-steepening \cite{moses:2006b,zhou:2012,Guo:2013}. Especially the self-steepening can become strong in cascading since it contributes to the intrinsic Kerr-induced self-steepening with a scaling factor $d_{12}/\Delta k$, i.e. the ratio between GVM and phase mismatch \cite{bache:2007a,moses:2006b,ilday:2004}. In the spectral domain, significant spectral broadening occurs up to the self-compression point, and after this stage a broadband dispersive wave forms in the range 4.0-$5.0\mic$. This peak can also be noticed in the experimental spectrum in Fig. \ref{fig:SCG}(a). Towards shorter wavelengths the second- and third-harmonic spectra are evident, although the latter is quite weak. Notice that the harmonic spectra also show continuum generation; this is because they are not phase-matched, and in the cascading limit the harmonic spectrum is "slaved"  or "locked" to the pump spectrum \cite{bache:2007a,Valiulis:2011,Zhou:2014a}. 

\begin{figure}[tb]
\includegraphics[width=\linewidth]{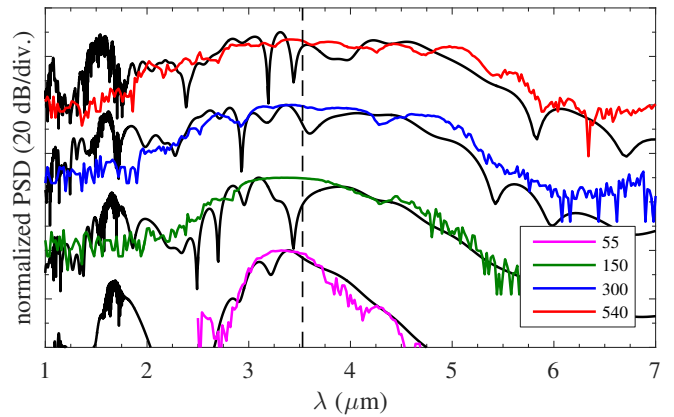}
\caption{\label{fig:exp-sim} 
Normalized PSD at the crystal exit, comparing numerical simulations and experimental results for $\lambda_1=3.39\mic$. A 30 dB offset per dataset was used to make the plot easier to read. The experimental intensities in GW/cm$^2$ are shown in the legend, and the numerical simulations (black lines) used similar intensities. }%
\end{figure}

Figure \ref{fig:exp-sim} compares simulations for the $\lambda_1=3.39\mic$ case qualitatively with the experimental data for different intensities. The simulations show a good agreement concerning the spectral bandwidth, especially in the mid-IR. In the near-IR the simulations have a clear peak at the SH wavelength (although it becomes somewhat blue-shifted at high intensities, an effect that is already apparent in Fig. \ref{fig:sim}(e) despite the moderate intensity used there), and the experimental data do not show this. This could be because the plane-wave simulations only model the on-axis distribution, while in the  experiment the aperture of the mid-IR monochromator was quite big so it essentially records a spatial average over most of the beam. Another explanation, at least for high intensities, is instead that diffraction can no longer be neglected, as we show below. Thus, the assumptions behind the plane-wave model breaks down and a full 3+1D simulation must be carried out. 

We also attempted, along the lines of the results shown in \cite{ashihara:2002,moses:2006,zhou:2012,Zhou:2015-OL}, to measure a self-compressed soliton with a home-built intensity autocorrelator. However, we were not able to find clear signs of a clean compressed soliton, and to understand this, the detailed numerical simulations in Fig. \ref{fig:sim} show multiple pulse splitting. This means that a single compressed pulse is rarely seen and instead two or more short pulses form, which are much harder to detect with an autocorrelation unit. 

The numerical simulations included noise to gauge the coherence of the spectra. The chosen noise source corresponds to the quantum fluctuations in the vacuum state according to the Wigner representation \cite{werner:1995,werner:1997,brambilla:2004}. The advantage of the Wigner representation (compared to P or Q) is that the noise is additive and is only added in the initial condition; this means that the standard split-step routine based on Runge-Kutta ODE solvers can be use. Conversely more rigorous Langevin solvers like the Heun method are needed for the multiplicative noise of the P and Q representation, i.e. where noise must be added at every $z$ step. In order to model the vacuum state in the Wigner representation, a half photon per temporal mode is added to the initial condition with Gaussian white noise distribution \cite{brambilla:2004}. This approach is different than the usual one-photon-per-mode approach \cite{Dudley2002}, which adds a Gaussian distributed phase noise in frequency domain with 1 photon per frequency mode average energy density. In time domain this approach has on average $2\pi$ half photons per mode; this does not conform with the vacuum fluctuations of the Wigner model. With this noise implementation, we repeated each simulation case a suitable number of times with different initial noise realizations to be able to calculate the first order spectral coherence function $g_{12}^{(1)}$, and we found that the coherence of the spectra to be excellent: an example is shown in Fig. \ref{fig:sim}(c) where the spetral coherence function is unity across the generated supercontinuum. We even attempted to add an additional significantly stronger noise source than the vacuum fluctuations (i.e. noise from the amplification stage or technical noise sources, all modeled in the same way as a Gaussian white noise but with many photons per temporal mode), and the high coherence still pertained.  %This is typical when femtosecond pumping is employed.

In the experiment we also characterized the transverse cross section of the output beam with an uncooled microbolometer camera. The camera is sensitive in the entire range of the recorded supercontinua. The measurements were done around 20 cm from the exit of the crystal, and from the beam waist size of the input beam ($w_0=0.2$ mm) we estimate that the distance $z/z_R\simeq 5$, i.e. 5 times the Rayleigh length. This places the camera in the Fresnel diffraction zone. The evolution of the beam vs. intensity showed formation of a more narrow spot in the center. Although this was not a near-field measurement, it does indicate diffraction for high intensities. This could be related to the pulse compression but it might also indicate that the self-defocusing cascading and self-focusing material Kerr effects are competing and give nontrivial spatio-temporal coupling. Further experiments are needed to clarify this. 

That being said, the spectra and the spatial characterization did not reveal any  fluctuations in the transverse beam profile on a shot-to-shot level and the long-term stability was good. Thus, despite the diffraction that we observed, the generated supercontinua are stable and should be quite useful for spectroscopy. As a case in point, we verified that the supercontinuum could be used to do simple absorption measurements: by passing the supercontinuum through an LN crystal, we could reconstruct transmission curves quite similar to FTIR measurements of the crystal (in particular the 4-$6\mic$ mid-IR absorption edge matched very well).

\section{Conclusion}

We have experimentally demonstrated multiple-octave spanning supercontinuum generation by pumping a bulk lithium thioindate (LIS) crystal in the 3.0-$3.9\mic$ range with bright 50 $\mu$J 85 fs pulses. The crystal was pumped with a loosely focused 0.27 mm diameter beam and the onset of a supercontinuum (octave spanning) was typically around 150 GW/cm$^2$, i.e. around 10 $\mu$J of pulse energy. 
The crystal was cut for noncritical phase-mismatched SHG, giving a strong self-defocusing nonlinearity. We did see diffraction effects in the transverse part of the beam at high intensities, and we will in future experiments look into the cause of this. This notwithstanding, the supercontinua at high intensities did reveal any significant spatial fluctuations and should therefore be quite useful for spectroscopy. 
The broadest spectrum spanned the range from 1.6-$7.0\mic$, enough to cover the entire range of fundamental vibrational frequencies of hydrogen and carbon-type bonds, including the near-IR part needed to cover their first overtones. Especially because the overtones can be probed as well, this makes it extremely desirable for mid-IR vibrational spectroscopy. This technique can readily be used in other mid-IR quadratic nonlinear crystals as well, which means that other parts of the mid-IR can be covered, especially the fingerprint region from 6-$12\mic$, by pumping similar crystals at longer wavelengths. 

\begin{acknowledgments}
This work has been supported by the Danish Council for Independent Research (grants 11-106702 and 4070-00114B). We thank Poul B. Petersen, Satoshi Ashihara, Valentin Petrov, Jens Biegert and Cord Arnold for fruitful discussions.
\end{acknowledgments}

%\bibstyle{aipnum4-1}
\bibliography{literature}
%\bibliography{M:/homeextra/Bibtex/literature}

\end{document}